\documentclass[
  reprint,
  aps,
  prl,
  amsmath,
  amssymb,
  superscriptaddress
]{revtex4-2}
\usepackage{graphicx}
\usepackage{dcolumn}
\usepackage{bm}
\usepackage{physics}
\usepackage{amsmath}
\usepackage[version=3]{mhchem}
\usepackage{subfig}
\usepackage{color}
\usepackage{ifthen}
\newcounter{num}

\newcommand{\fix}[1]{#1}

\newcommand{\vk}{\vb*{k}}
\newcommand{\vkp}{\vb*{k}_{+}}
\newcommand{\vkm}{\vb*{k}_{-}}
\newcommand{\vq}{\vb*{q}}
\newcommand{\vqz}{\vb*{q}_{0}}
\newcommand{\vp}{\vb*{p}}
\newcommand{\vpp}{\vb*{p}_{+}}
\newcommand{\vpm}{\vb*{p}_{-}}
\renewcommand{\vr}{\vb*{r}}

\newcommand{\gre}{\mathcal{G}}
\newcommand{\grep}{\mathcal{G}_{+}}

\newcommand{\mean}[1]{\left\langle #1 \right\rangle}
\newcommand{\ep}{\varepsilon}
\newcommand{\epk}{\ep_{\vk}}

\newcommand{\ham}{\hat{\mathcal{H}}}

\newcommand{\pha}[1][]{\hat{\alpha}_{#1}}
\newcommand{\phb}[1][]{\hat{\beta}_{#1}}
\newcommand{\phC}[2]{\mathcal{C}\pqty{ #1 , #2 }}
\newcommand{\phD}[2]{\mathcal{D}\pqty{ #1 , #2 }}

\newcommand{\iwlam}{i\omega_{\lambda}}

\newcommand{\iwm}{i\omega_{m}}
\newcommand{\iwmp}{i\omega_{m+}}
\newcommand{\ien}{i\varepsilon_{n}}
\newcommand{\ienp}{i\varepsilon_{n+}}

\newcommand{\gp}{g_{\vp}}
\newcommand{\gpp}{g_{\vpp}}
\newcommand{\gpm}{g_{\vpm}}

\newcommand{\Trb}[1]{\mathrm{Tr}\left\lbrack #1 \right\rbrack}

\newcommand{\sumkpnm}{\sum_{\vk,\vp,n,m}}
\newcommand{\sumkpnma}{\sum_{\vk,\vp,n,m,\alpha}}

\newcommand{\vkbttwo}{\frac{(k_{B}T)^2}{V^2}}
\newcommand{\kbt}{k_{B}T}
\newcommand{\coeff}[1]{
\ifthenelse{\equal{#1}{1}}{\frac{ie^{2}\hbar c_{L}^{2} p_{x}}{2m^{*}}}\relax
\ifthenelse{\equal{#1}{2}}{\frac{ie^{2}c_{L}^{2} p_{x}}{2\hbar}}\relax
\ifthenelse{\equal{#1}{3}}{\frac{ie^{2}}{m^{*}\hbar}}\relax
\ifthenelse{\equal{#1}{4}}{\frac{ie^{2}}{\hbar^{3}}}\relax
\ifthenelse{\equal{#1}{5}}{\frac{ie^{2}c_{L} p_{x}}{m^{*}p}}\relax
\ifthenelse{\equal{#1}{6}}{\frac{ie^{2}c_{L} p_{x}}{\hbar^{2}p}}\relax
}
\newcommand{\coeffb}[1]{
\ifthenelse{\equal{#1}{1}}{\frac{ie^{2}B_{z}\hbar c_{L}^{2} p_{x}}{4m^{*}}}\relax
\ifthenelse{\equal{#1}{2}}{\frac{ie^{2}B_{z} c_{L}^{2} p_{x}}{4\hbar}}\relax
\ifthenelse{\equal{#1}{3}}{\frac{ie^{2}B_{z}}{2m^{*}\hbar}}\relax
\ifthenelse{\equal{#1}{4}}{\frac{ie^{2}B_{z}}{2\hbar^{3}}}\relax
\ifthenelse{\equal{#1}{5}}{\frac{ie^{2}B_{z} c_{L} p_{x}}{2m^{*}p}}\relax
\ifthenelse{\equal{#1}{6}}{\frac{ie^{2}B_{z} c_{L} p_{x}}{2\hbar^{2}p}}\relax
}
\newcommand{\coefff}[1]{
\ifthenelse{\equal{#1}{1}}{\frac{e^{2}B_{z}\hbar^{4}\omega c_{L}^{2} g'^{2}}{4m^{*~2}\varGamma_{ph}}}\relax
}
\allowdisplaybreaks

\newif\ifshowmain
\showmaintrue

\newif\ifshowsupp
\showsupptrue

\begin{document}

\ifshowmain
    \title{Phonon Drag Effect in Nernst and Thermal Hall Effects: General Theory and Application to Dilute Metal SrTiO$_{3-\delta}$}

    \author{\normalsize Junya Endo}
    \affiliation{Department of Physics, University of Tokyo, Bunkyo, Tokyo 113-0033, Japan}

    \author{\normalsize Hiroyasu Matsuura}%
    \affiliation{Department of Physics, University of Tokyo, Bunkyo, Tokyo 113-0033, Japan}

    \author{\normalsize Masao Ogata}
    \affiliation{Department of Physics, University of Tokyo, Bunkyo, Tokyo 113-0033, Japan}
    \affiliation{Trans-scale Quantum Science Institute, University of Tokyo, Bunkyo-ku, Tokyo 113-0033, Japan}

    \date{\today}

    \begin{abstract}
        In magnetic fields, thermal gradient-induced effects such as the Nernst and thermal Hall effects are significantly influenced by phonon drag, which works in conjunction with the force on electrons \fix{in a magnetic field}.
We introduce a method to calculate Nernst and thermal Hall conductivities influenced by phonon drag using linear response theory to treat the magnetic field as a first-order perturbation.
Our formula is general enough to apply to \fix{various systems in which} the Green's functions of electrons and phonons are given.
We apply the obtained general theory to the recent experiments of dilute metal SrTiO$_{3-\delta}$, known for strong Nernst and thermal Hall effects due to phonon drag.
We find good agreement even quantitatively.
This is notable as all model parameters are derived from experimental data without adjustable parameters.

    \end{abstract}

    \maketitle

    \paragraph{Introduction.}
    \label{section:introduction}
    The thermoelectric effect, which converts temperature gradients into voltage, has recently been attracting much attention as environmentally friendly power generation method that achieves carbon neutrality\cite{Bell,Koumoto,Petsagkourakis,Liu,Hendricks}.
To improve the efficiency of thermoelectric materials for electricity generation, it is crucial to understand the underlying mechanisms that enhance the thermoelectric effect.

Phonon drag, a momentum transfer between electrons and phonons, is important to enhance the thermoelectric effect at low temperatures where the phonons have long life time.
First suggested by Gurevich in the 1940s\cite{Gurevich} and later confirmed in semiconductors\cite{Frederikse,Geballe1954,Geballe1955}, significant theoretical developments have been made particularly using the Boltzmann equation\cite{Herring,Ziman}.
Recently huge Seebeck effect due to phonon drag effect observed in FeSb$_{2}$\cite{Bentien,Sun,Takahashi2016} has been analyzed with microscopic theory\cite{Matsuura} employing Kubo and Luttinger's linear response theory\cite{Kubo,Luttinger}.
This analysis has provided a deeper insight into the thermoelectric effect closely connected to the impurity band.
The growing use of microscopic theory is paving the way for further research into thermoelectricity.

Recently, SrTiO$_{3-\delta}$, characterized by a small Fermi surface and metallic conductivity at low temperatures, demonstrates huge Seebeck, Nernst, and thermal Hall effects at low temperatures \fix{around $20~\mathrm{K}$ \cite{Cain, Behnia_Seebeck, Behnia}}.
The temperature at which these quantities become maximum coincides with those at which the thermal conductivity, mainly due to phonons, becomes maximum.
\fix{These experimental results suggest that the phonon drag effect dominates the huge Seebeck, Nernst, and thermal Hall effects at around $20~\mathrm{K}$.}
This issue is fundamentally and microscopically important because phonon drag arises from the interaction between two of the most essential components of matter: electrons and phonons.
However, previous theories on transverse responses involving the phonon drag have been limited to the phenomenological method based on the Boltzmann equation\cite{Price, Miele}, and a comprehensive microscopic theory has been lacking.
The challenge lies in the fact that \fix{dealing with the force on electrons in magnetic field is} exceedingly complex to address in quantum mechanics.

In this letter, we present a general microscopic theory that addresses both longitudinal and transverse thermal responses incorporating phonon drag and interband effects of a magnetic field based on Kubo and Luttinger's linear response theory.

Since our microscopic theory is quite general, it can be applied to various cases to investigate the possible enhancement of thermoelectric properties.
In addition, by applying the obtained theory, we derive a compact formula for a simple case of a single band.
Finally, we show that the experimental results reported in SrTiO$_{3-\delta}$ are reproduced by the compact formula even quantitatively.

    \paragraph{Method.}
    \label{section:method}
    The electrical current density $\vb*{J}^{1}$ and the thermal current density $\vb*{J}^{2}$ are given in the first-order responses of the external electric field $\vb*{E}$ and temperature gradient $\nabla T$ as follows\cite{Mahan}:
\begin{align}
    \vb*{J}^{1} & = \overline{L}^{11} \vb*{E} + \overline{L}^{12} \pqty{ - \frac{\nabla T}{T}}, \\
    \vb*{J}^{2} & = \overline{L}^{21} \vb*{E} + \overline{L}^{22} \pqty{ - \frac{\nabla T}{T}},
\end{align}
where $\overline{L}$ is a linear response coefficient.
When a magnetic field is applied along the $z$ direction, the off-diagonal components of the coefficients ($L_{xy}^{ij},~i,j=1,2$) can be finite, and thus a 2 $\times$ 2 matrix of linear response coefficients should be considered when discussing the transport properties in the $xy$ plane.
The Hamiltonian density studied in this letter is given as\cite{Mahan}
\begin{align}
    \hat{h}(\vr)                & = \hat{h}^{\mathrm{ele}}(\vr) + \hat{h}^{\mathrm{ph}}(\vr) + \hat{h}^{\mathrm{e-p}}(\vr),                         \\
    \hat{h}^{\mathrm{ele}}(\vr) & = \frac{1}{2m}\pqty{\hat{\vp} - e\vb*{A}(\vr)}^{2} + V(\vr),
    \\
    \hat{h}^{\mathrm{ph}}(\vr)  & = \frac{\hat{\vb*{P}}(\vr)^{2}}{2M} + \frac{Mc_{L}^{2}}{2}\sum_{\mu}\pqty{\pdv{\hat{\vb*{u}}(\vr)}{r_{\mu}}}^{2}, \\
    \label{eq:Hamiltonian_ph}
    \hat{h}^{\mathrm{e-p}}(\vr) & = -g\hat{\rho}(\vr)\sum_{\mu}\pqty{\pdv{u_{\mu}(\vr)}{r_{\mu}}},
\end{align}
where \fix{$\mu = x,y,z$, and} ${h}^{\mathrm{ele}}(\vr)$, $\hat{h}^{\mathrm{ph}}(\vr)$, and $\hat{h}^{\mathrm{e-p}}(\vr)$ are Hamiltonian densities of electrons, phonons, and electron-phonon coupling, respectively.
Here $V(\vr)$ is periodic potential for electrons leading to the Bloch bands, $\hat{\vb*{u}}(\vr)$ and $\hat{\vb*{P}}(\vr)$ are lattice displacement operator and its momentum, $M$ is the mass of atoms, $c_{L}$ is the phonon velocity, $\hat{\rho}(\vr)$ is the electron density operator, and $g$ is the electron-phonon coupling constant.
\fix{In this letter, we focus on longitudinal acoustic phonons, as transverse modes do not interact with electrons in the lowest order with respect to the amplitude of phonons. Optical phonons are not considered, as their high energy renders them irrelevant for this study.}
In this Hamiltonian, the linear response coefficients $L_{xx}^{ij}$ and $L_{xy}^{ij}$ due to the phonon drag is calculated by Kubo-Luttinger's linear response theory\cite{Kubo,Luttinger} as
\begin{equation}
    L^{ij}_{\mu\nu} = \lim_{\omega \to +0} \frac{\Phi^{ij}_{\mu\nu}\pqty{\vb*{0},\hbar(\omega + i\delta)} - \Phi^{ij}_{\mu\nu}\pqty{\vb*{0}, i\hbar \delta}}{i(\omega + i\delta)}.
\end{equation}
Here, the correlation function $\Phi^{ij}_{\mu\nu}(\vq,\iwlam)$ is defined as
\begin{equation}
    \Phi^{ij}_{\mu\nu}(\vq,\iwlam) = \frac{1}{V}\int_{0}^{\beta}\dd \tau ~ \mean{j^{i}_{\vq\mu}(\tau) j^{j}_{\vb*{0}\nu}(0)} \mathrm{e}^{\iwlam \tau},
\end{equation}
where $\vb*{j}^{1}_{\vq}(\tau)$ and $\vb*{j}^{2}_{\vq}(\tau)$ are the Fourier transform of the current and thermal current density operators in the Heisenberg representation, respectively, and $\iwlam$ represents the frequency of the external field (in this case, electric field).
Since the $\vr$-dependence of the vector potential $\vb*{A}(\vr)$ is difficult to treat, we use the method to introduce the Fourier component $\vb*{A}(\vr) = - i\vb*{A}_{\vqz} \mathrm{e}^{i\vqz\cdot\vr}$, as studied by Hebborn et al. and by Fukuyama\cite{Hebborn, Fukuyama1971}.
Then we calculate $\Phi_{\mu\nu}^{ij}(\vqz, \iwlam)$ in the linear order with respect to $\vqz$ and $\vb*{A}_{\vqz}$.
To guarantee the gauge invariance, the physical quantity should be proportional to $q_{0,x}A_{\vqz, y} - q_{0,y}A_{\vqz, x} = B_{z}$, and at the end, we take the limit of $\vqz \to \vb*{0}$.
As for the electronic state, we use the Luttinger-Kohn (LK) representation\cite{LK,Fukuyama1969}.
\par
Without magnetic field, which was studied before\cite{Baumann, Ogata}, we obtain
\begin{equation}
    \begin{split}
         & \Phi_{xx}^{\mathrm{PD}}(\vb*{0}, \iwlam)
        = \frac{(\kbt)^{2}}{V^{2}} \sumkpnm e^2\hbar c_{L}^{2}p_{x} \gp^{2} \phC{\vp}{\iwm}          \\
         & \times\phD{\vp}{\iwmp}\Trb{\gre\gamma_{\vk,x}\grep\gre' + \gre\gamma_{\vk,x}\grep\grep'},
    \end{split}
\end{equation}
where \fix{$g_{\vp}$ is related to the electron-phonon coupling $g$ as} $g_{\vp} = g \sqrt{\hbar\abs{\vp} / 2Mc_{L}}$ \cite{Mahan, Supplemental}, $\ien (\iwm)$ is the fermionic (bosonic) Matsubara frequency, $\gamma_{\vk,\mu}= \partial\ham/\partial k_{\mu}$, $\gre = \gre(\vk,\ien)$ is an electron thermal Green's function, and $\grep = \gre(\vk,\ienp)$, $\gre' = \gre(\vk-\vp,\ien-\iwm)$ with $\ienp = \ien + \iwlam$ and $\iwmp = \iwm + \iwlam$. \fix{Here, $\gre$ and $\gamma_{\vk,\mu}$ are the matrix representation whose basis are the Bloch bands, and $\mathrm{Tr}$ represents the trace over the band index including the spin degrees of freedom.}
$\phC{\vp}{\iwm}$ and $\phD{\vp}{\iwm}$ are two kinds of phonon thermal Green's functions
\fix{defined as
    \begin{align}
        \phD{\vp}{\tau - \tau'} & = - \mean{\mathrm{T}_{\tau} \pha[\vp]\pqty{\tau}\pha[-\vp]\pqty{\tau'}}, \\
        \phC{\vp}{\tau - \tau'} & = - \mean{\mathrm{T}_{\tau} \pha[\vp]\pqty{\tau}\phb[\vp]\pqty{\tau'}},
    \end{align}
    where $\pha[\vp] = \hat{a}_{\vp} + \hat{a}^{\dagger}_{-\vp}$ and $\phb[\vp] = -\hat{a}_{-\vp} + \hat{a}^{\dagger}_{\vp}$ with $\hat{a}_{\vp}~(\hat{a}^{\dagger}_{\vp})$ being the phonon annihilation (creation) operator.}
\par
\begin{figure}[htbp]
    \begin{center}
        \includegraphics[keepaspectratio,width=0.65\linewidth]{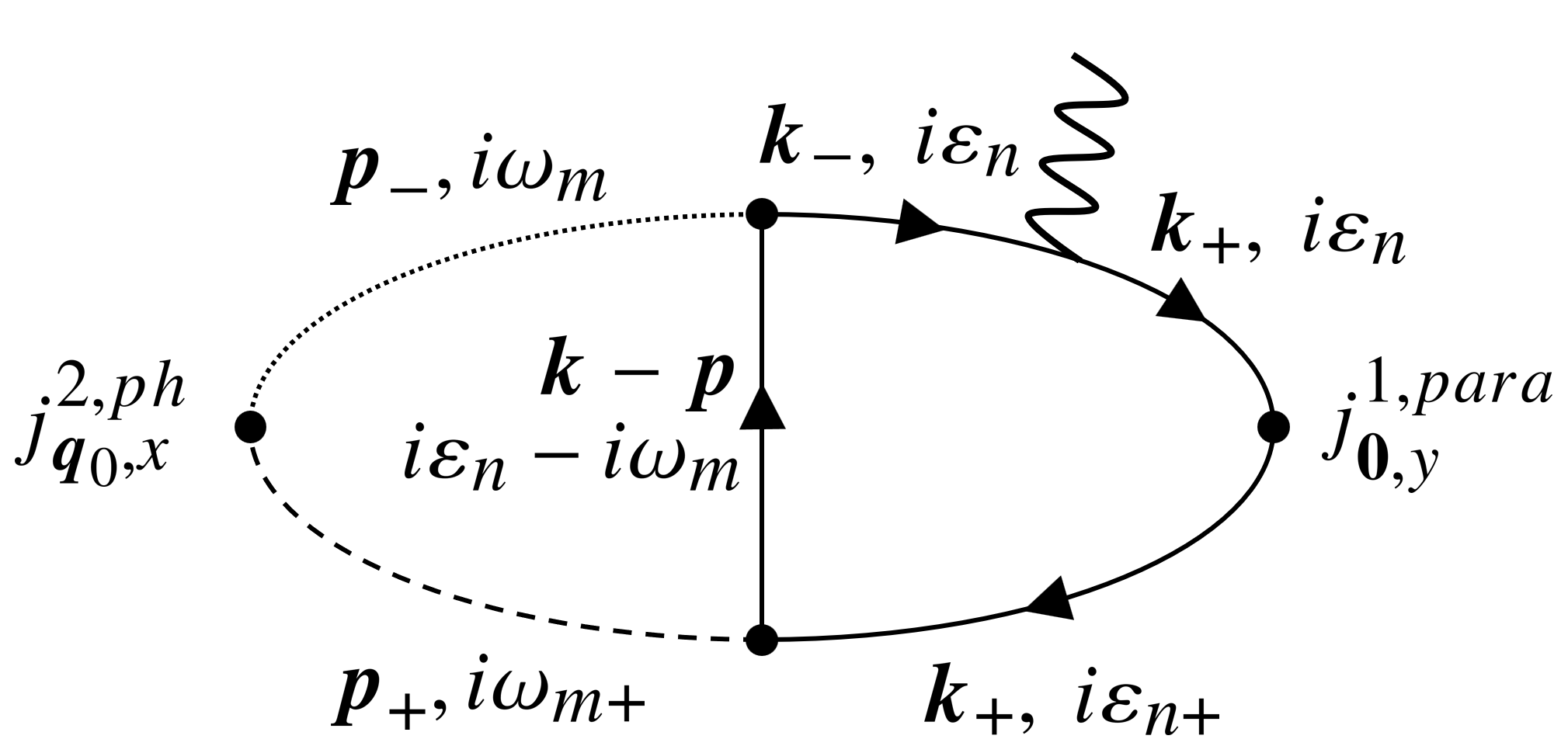}
        \caption{A typical Feynman diagram corresponding to $\Phi^{21,\mathrm{PD}}_{xy}$ of the phonon drag effect. Solid lines, dotted lines, dashed lines, and a wavy line correspond to electron Green's functions, phonon Green's functions $\mathcal{C}$, $\mathcal{D}$, and vector potential due to magnetic field, respectively.
        The left-hand-side vertex is the thermal current of phonons $\vb*{j}^{2,\mathrm{ph}}$ due to phonons and the right-hand-side vertex is the electric current $\vb*{j}^{1}$.
        Here $\vk_{\pm} = \vk \pm \vqz / 2$, and $\vp_{\pm} = \vp \pm \vqz / 2$.
        The all diagrams are shown in Supplemental Material\cite{Supplemental}.
        }
        \label{fig:Typical_diagram}
    \end{center}
\end{figure}
\fix{To calculate the phonon-drag contribution to $\Phi^{21,\mathrm{PD}}_{xy}$ in the magnetic field,} there are eight kinds of Feynman diagrams with different momentum and thermal frequency assignments (see Fig. S1 in Supplemental Material \cite{Supplemental}).
\fix{A typical Feynman diagram of $\Phi^{21,\mathrm{PD}}_{xy}$ is shown in Fig. \ref{fig:Typical_diagram}, where we choose the thermal current $\vb*{j}^{2,\mathrm{ph}}$ for the left-hand-side vertex and the electric current $\vb*{j}^{1}$ for the right-hand-side vertex. The effect of the vector potential $\vb*{A}_{\vqz}$ is represented by a wavy line.}
Summing up all the diagrams and checking the gauge invariance, we obtain
\begin{widetext}
    \begin{equation} \label{eq:general_result}
        \begin{split}
              & \Phi_{xy}^{21,\mathrm{PD}}(\vqz\rightarrow \vb*{0}, \iwlam)            \\
            = & \vkbttwo \sumkpnm \coeffb{1} \gp^{2} \phC{\vp}{\iwm}\phD{\vp}{\iwmp}
            \Trb{\gre\gre'\grep\gamma_{\vk,x}\grep - \gre\gamma_{\vk,x}\gre\gre'\grep +
            (\vb*{p} \rightarrow -\vb*{p}, \iwm \rightarrow -\iwm-i\omega_\lambda) }   \\
              & + \vkbttwo \sumkpnm \coeffb{2} \gp^{2} \phC{\vp}{\iwm}\phD{\vp}{\iwmp}
            \Trb{ ( \gamma_{\vk,y}\gre\gamma_{\vk,x}\gre\gamma_{\vk,y}\gre\gre'\grep
            - \gamma_{\vk,y}\gre\gamma_{\vk,y}\gre\gamma_{\vk,x}\gre\gre'\grep \right. \\
              & \quad+
                \gamma_{\vk,y}\gre\gre'\grep\gamma_{\vk,x}\grep\gamma_{\vk,y}\grep
                - \gamma_{\vk,y}\gre\gre'\grep\gamma_{\vk,y}\grep\gamma_{\vk,x}\grep +
                \gamma_{\vk,y}\gre\gamma_{\vk,y}\gre\gre'\grep\gamma_{\vk,x}\grep
            - \gamma_{\vk,y}\gre\gamma_{\vk,x}\gre\gre'\grep\gamma_{\vk,y}\grep        \\ &\quad+
                \gamma_{\vk,y}\gre\gre'\gamma_{\vk-\vp,x}\gre'\gamma_{\vk-\vp,y}\gre'\grep
                - \gamma_{\vk,y}\gre\gre'\gamma_{\vk-\vp,y}\gre'\gamma_{\vk-\vp,x}\gre'\grep )
                +  (\vb*{p} \rightarrow -\vb*{p}, \iwm \rightarrow -\iwm-i\omega_\lambda)
                \left.  },
        \end{split}
    \end{equation}
\end{widetext}
where $(\vb*{p} \rightarrow -\vb*{p}, \iwm \rightarrow -\iwm-i\omega_\lambda)$ means the contributions with the replacements  $\vb*{p} \rightarrow -\vb*{p}$ and
$\iwm \rightarrow -\iwm-i\omega_\lambda$ in the preceding terms.
\par
The above expression is one of the main results of the present letter, which gives the general form of the phonon drag contribution in $L^{21}_{xy}$ in the linear-order of the magnetic field.
\fix{The Green's functions can be permitted to be multiband, and the obtained result is applicable to multiband systems. Note that we have not taken into account vertex corrections.}
For the thermal Hall effect, it is easy to see that $\Phi_{xy}^{22,\mathrm{PD}}$ is given by multiplying the term inside the summation of Eq. \eqref{eq:general_result} by $\ep_{\vk} - \mu$.

    \paragraph{Case of Free Electrons.}
    \label{section:free}
    In the following, we study the case in which the electron system is described as a single band with a dispersion $\ep_{\vk} = \hbar^{2}k^{2} / 2m^{*}$, with $m^{*}$ being an effective mass.
For simplicity, the relaxation rate of electrons $\varGamma ( = \hbar / 2\tau)$ is assumed to be independent of temperature.
We assume that the real part of the self-energy is renormalized into the chemical potential and $m^{*}$.
In this case, the Green's function of the electron is $\gre(\vk, \ien) = \pqty{\ien - \ep_{\vk} + \mu + i\varGamma\mathrm{sign}(\ep_{n})}^{-1}$.
Let the relaxation rate of the phonon be written as $\varGamma_{\mathrm{ph}}(\omega)$, where $\omega$ is the phonon energy.
Furthermore, we assume that the phonon relaxation rate is small ($\varGamma_{\mathrm{ph}}(\omega) \ll \ep_{F}$), and we pick up the leading order contributions which is in the order of $\ep_{F} / \varGamma_{\mathrm{ph}}(\omega)$.
In this case, the sum of the Matsubara frequencies and the sum of the wavenumbers in Eq. \eqref{eq:general_result} can be performed analytically, and we obtain the following compact expression for $L_{xy}^{21}$ after a straightforward calculation (details of derivation is shown in Supplementary Material\cite{Supplemental}):
\begin{equation}\label{eq:L_of_free}
    \begin{split}
        \frac{L_{xy}^{21, \mathrm{PD}}}{B_{z}} & =  -\frac{e^{2}g'^{2}}{48\pi^{3}\hbar^{5}c_{L}^3\varGamma^{2}}\int_{m^*c_L^2/2}^{\infty}\dd \ep \int_{0}^{\alpha} \dd \omega \\
                                               & \times \frac{\omega^{4}}{\varGamma_{\mathrm{ph}}(\omega)}\pqty{f(\ep - \omega) - f(\ep)}\pdv{n(\omega)}{\omega},
    \end{split}
\end{equation}
where $\alpha = 2\sqrt{2m^{*}c_{L}^{2} \ep} - 2m^{*}c_{L}^{2}$, which originates from the momentum and energy conservation, and $f(\ep)$ and $n(\omega)$ are the Fermi and Bose distribution functions, respectively.
\fix{Here, for convenience, we introduce a parameter $g'$ relating to electron-phonon coupling defined as $g' = g_{\vp} / \sqrt{\abs{\vp}}$.}
$L_{xy}^{22, \mathrm{PD}}$ is obtained by multiplying the integrand by $(\ep - \mu)/e$.
\par
Let us briefly discuss the dependence on the effective mass of $L^{21,\mathrm{PD}}_{xy}$.
A larger effective mass cause greater $L^{21,\mathrm{PD}}_{xy}$ when $m^{*}c_{L}^{2}, \ep_{F} \lesssim k_{B}T$, which is clarified numerically with neglecting the $\omega$ dependence of $\varGamma_{\mathrm{ph}}(\omega)$.
In that region, the thermoelectric coefficient is approximately proportional to the effective mass\cite{Supplemental}.
This is the case also for $L^{22,\mathrm{PD}}_{xy}$.
Note that the significance of large effective mass was also recognized in the context of the Seebeck effect\cite{Matsuura}.

    \paragraph{Application to SrTiO$_{3-\delta}$.}
    \label{section:applcation}
    \begin{figure}[tbp]
    \begin{center}
        \includegraphics[keepaspectratio, width=0.9\linewidth]{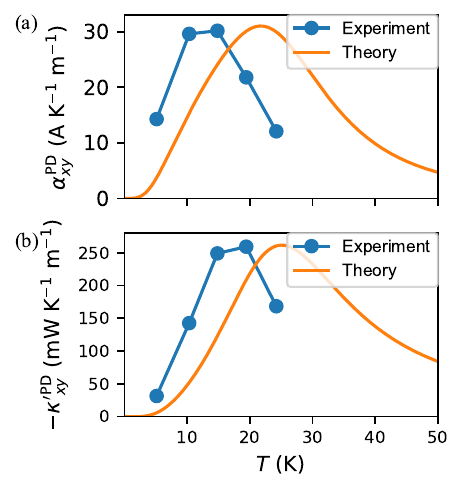}
        \caption{
            (a) Nernst conductivity and (b) thermal Hall conductivity due to phonon drag in a dilute metal at $B_{z} = 1~\mathrm{T}$. The orange lines indicate the theoretical results.
            The blue dots are the experimental results of SrTiO$_{3-\delta}$ obtained from \cite{Behnia}.}
        \label{fig:results}
    \end{center}
\end{figure}
\begin{figure}[tbp]
    \begin{center}
        \includegraphics[keepaspectratio, width=0.85\linewidth]{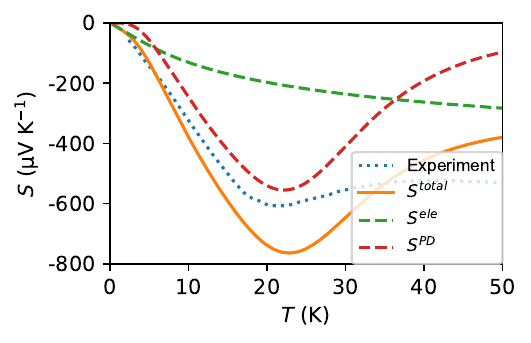}
        \caption{
            Temperature dependence of Seebeck coefficient.
            The red and green dashed lines are the Seebeck coefficient due to the phonon drag effect and electrons, respectively.
            The blue dotted line represents the experimental result of SrTiO$_{3-\delta}$ obtained from \cite{Behnia}.
        }
        \label{fig:Seebeck}
    \end{center}
\end{figure}
Next, we apply Eq. \eqref{eq:L_of_free} to a dilute metal SrTiO$_{3-\delta}$, which shows the peculiar thermal responses probably due to the phonon drag effect\cite{Behnia}.
For SrTiO$_{3-\delta}$, all the parameters in Eq. \eqref{eq:L_of_free} have been well studied and the actual values are known: $n = 1.0 \times 10^{18} ~\mathrm{cm}^{-3}$, $m^{*} = 1.8~m_{e}$\cite{Behnia2013,Behnia}, and $c_{L} = 7800~\mathrm{m/s}$\cite{Rehwald,Behnia}.

\fix{By deriving the  electron-phonon coupling from deformation potential\cite{Abrikosov}, we obtain $g' = \sqrt{\hbar E_{d}^{2} / 2 \rho_{M} c_{L}}$. Its value} is calculated as $g' \simeq 4.6 \times 10^{-21}~\mathrm{eV~m^{2}}$ using a deformation potential $E_{d} = 4.0~\mathrm{eV}$\cite{Janotti, Verma} and a mass density $\rho_{M} = 5120~\mathrm{kg~m^{-3}}$\cite{Verma}.
\fix{It is widely recognized that there are soft modes present in SrTiO$_{3-\delta}$\cite{Cochran,Anderson,Ngai}. However, because these soft modes are transverse modes, we do not consider their effects as discussed below Eq. \eqref{eq:Hamiltonian_ph}.}

To estimate $\varGamma$ for electrons, we use the resistivity of SrTiO$_{3-\delta}$.
Although the resistivity depends on temperature, we use the value $\rho \simeq 1.4 \times 10^{-5}~\mathrm{\Omega~m}$\cite{Behnia} around $20~\mathrm{K}$ since we focus on the low temperature region where the thermoelectric coefficient has a peak. Using the above value of $\rho$, we obtain $\varGamma = \hbar e^{2} \rho n / 2 m^{*} \simeq 7 \times 10^{-5}~\mathrm{eV}$.
The chemical potential is determined self-consistently from $m^{*}$ and $n$\cite{Supplemental}, which gives the Fermi energy and the Fermi wavenumber as  approximately $2.0~\mathrm{meV}$ and $0.031~\mathrm{\AA}^{-1}$, respectively.
In the following, we assume $\varGamma_{\mathrm{ph}}$ is a function of $\omega = \hbar\omega_{\vp}$ and $T$.
\fix{Since the microscopic calculation of the phonon relaxation is difficult and it is beyond the scope of this letter, we assume $\varGamma_{\mathrm{ph}}$ has the form\cite{Holland}
    \begin{equation}
        \label{eq:Gamma_ph}
        \varGamma_{\mathrm{ph}}(\omega) = \varGamma_{\mathrm{ph}}^{(0)} \pqty{1 + A\omega^{4}+B\kbt\omega^{2}\mathrm{e}^{-\hbar\omega_{D}/3\kbt}},
    \end{equation}
    where $\omega_{D}$ is the Debye frequency, for which we assume $\hbar \omega_{D} / k_{B} = 400~\mathrm{K}$ as a typical value. The coefficients, $\varGamma_{\mathrm{ph}}^{(0)}$, $A$, and $B$, are determined so as to reproduce the experimental data of longitudinal thermal conductivity $\kappa_{xx}$ shown in Ref. \cite{Behnia}. Then, we obtain $\varGamma_{\mathrm{ph}}^{(0)} \simeq 4.7 \times 10^{-7}~\mathrm{eV}$, $A \simeq 1.2 \times 10^{7}~\mathrm{eV^{-4}}$, and $B \simeq 1.7 \times 10^{9}~\mathrm{eV^{-2}}$\cite{Supplemental}.
}
\par
The obtained $\alpha^{\mathrm{PD}}_{xy} = L^{21,\mathrm{PD}}_{xy} / T$ at $B_{z} = 1~\mathrm{T}$ is shown as a solid line in Fig. \ref{fig:results}(a).
We can see that $\alpha_{xy}^{\mathrm{PD}}$ has a peak at around $20~\mathrm{K}$.
Note that the decrease of $\alpha^{\mathrm{PD}}_{xy}$ above $20~\mathrm{K}$ is mainly due to the enhancement of $\varGamma_{\mathrm{ph}}$.
The corresponding experimental results at $B_{z} = 1~\mathrm{T}$ are also indicated by blue dots in Fig. \ref{fig:results}(a)\cite{Behnia}.
The calculated $\alpha^{\mathrm{PD}}_{xy}$ is in good agreement with the experimental results considering that the present theory does not have any fitting parameters.
There is a slight difference between the experimental and theoretical peak temperatures, which will originate from the ambiguity of $\varGamma_{\mathrm{ph}}$ estimated from $\kappa_{xx}$, for example, due to the effect of optical phonons.
\par
Next, we discuss $\kappa'^{\mathrm{PD}}_{xy} = L^{22,\mathrm{PD}}_{xy} / T$ shown in Fig. \ref{fig:results}(b) compared with the experimental results\cite{Behnia}.
Again, the present theory explains experiments remarkably well.
Note that experimental measurements of $\kappa'$ include contributions other than phonon drag.
However, according to the analysis of Ref.\cite{Behnia}, the other contributions to $\kappa$ are negligibly small in the temperature range in Fig. \ref{fig:results}(b).
\par
One of the main reasons for the large values of $\alpha^{\mathrm{PD}}_{xy}$ and $\kappa'^{\mathrm{PD}}_{xy}$ is that the Fermi energy is small.
In fact, it can be numerically confirmed that large coefficients are obtained for small Fermi energy of a few meV.
Supplementary shows the Fermi energy dependence of these coefficients\cite{Supplemental}.
\par
To confirm the consistency in the present theory, we study phonon drag effect of Seebeck coefficient $S = L^{12}_{xx}/TL^{11}_{xx}$ without magnetic field.
Figure \ref{fig:Seebeck} shows the experimental and theoretical results of the temperature dependence of the Seebeck coefficient.
The experimental results (blue dotted line) show a peak at around $20~\mathrm{K}$, suggesting the presence of phonon drag\cite{Behnia}.
The green and red dashed lines ($S^{\mathrm{ele}}$ and $S^{\mathrm{PD}}$) indicate the theoretical results of $S^{\mathrm{ele}}= L^{12,\mathrm{ele}}_{xx}/TL^{11}_{xx}$ and $S^{\mathrm{PD}}= L^{12,\mathrm{PD}}_{xx}/TL^{11}_{xx}$.
Here, we use the linear response coefficients, $L^{11}_{xx}$, $L^{12,\mathrm{ele}}_{xx}$, and $L_{xx}^{12,\mathrm{PD}}$ obtained in \cite{Baumann,Ogata,Supplemental}.
The total Seebeck coefficient $S^{\mathrm{tot}} = S^{\mathrm{ele}} + S^{\mathrm{PD}}$(orange solid line) shows a peak at around $20~\mathrm{K}$, supporting the claims of the experimental evidence for phonon drag.
Note that $S^{\mathrm{ele}}$ is much larger than that of ordinary metals, which can be understood by considering $S^{\mathrm{ele}} \simeq \frac{\pi^{2} k_{B}^{2} T}{6e\ep_{F}}$ with the Fermi energy being small compared to the ordinary metals.

    \paragraph{Discussion.}
    \label{section:discussionn}
    \begin{figure}[tbp]
    \begin{center}
        \includegraphics[keepaspectratio,width=0.9\linewidth]{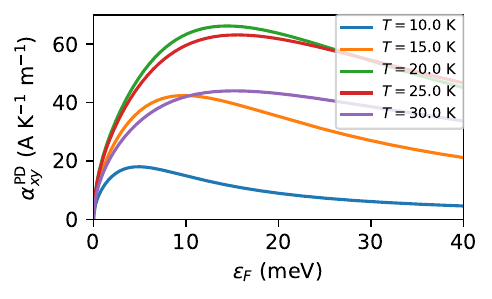}
        \caption{
            Fermi energy dependence of the Nernst conductivity.
            The values are obtained by hypothetically varying the Fermi energy of SrTiO$_{3-\delta}$.
            In the calculation, $\varGamma$ is assumed to be proportional to the Fermi wavenumber $k_{F}$ , and the other parameters of SrTiO$_{3-\delta}$ other than the chemical potential $\mu$ are fixed or independent of the Fermi energy.
        }
        \label{fig:Nernst_Conductivity}
    \end{center}
\end{figure}
Here, we compare ordinary metals, characterized by large Fermi energy $(\ep_{F} \gg \kbt, m^{*}c_{L}^{2})$, with dilute metals, examining both electron and phonon perspectives.
Let us discuss how transport coefficients depend on $\ep_{F}$.
Starting with electrons, we assume electron scattering is solely due to impurities, and thus proportional to the Fermi wavenumber $(\varGamma \propto k_{F})$\cite{AGD}.
As shown in Fig. \ref{fig:Nernst_Conductivity}, $\alpha_{xy}^{PD}$ increases as the Fermi energy decreases and is maximized for a Fermi energy of a few $\mathrm{meV}$.
(Seebeck coefficients and thermal Hall conductivities are shown in Supplementary\cite{Supplemental}.)
This can be qualitatively explained as follows: the phonon energy that contributes significantly to the transport coefficient is about $\omega \sim \kbt$, which can be seen from the integrand in Eq. \eqref{eq:L_of_free}.
On the other hand, from energy and momentum conservation, the electron and phonon energies, $\ep$ and $\omega$, involved in the phonon drag must satisfy $0 < \omega < \alpha = 2\sqrt{2m^{*}c_{L}^{2}\ep} - 2m^{*}c_{L}^{2}$.
Therefore, the magnitude of the transport coefficients is affected by whether $\ep_{F}$ is greater than $\sim (\kbt + m^{*}c_{L}^{2})^{2}/m^{*}c_{L}^{2}$.
For Fermi energies smaller than this, the transport coefficients are very small.
Given these behaviors, at least when $\varGamma_{\mathrm{ph}}(\omega)$ is invariant, a Fermi energy of about a few $\mathrm{meV}$ results in the largest transport coefficients.
\par
Turning to phonons:
In metals with large Fermi energies, electrons, due to their large momentum, primarily lose their momentum via electron-phonon interactions (Umklapp scattering).
However, in the dilute metals with low Fermi energy, Umklapp scattering due to the electron-phonon interaction is negligible, and momentum loss primarily occurs within
the phonon system.
This situation aligns with Herring's theory\cite{Herring}, which is generally applied to doped semiconductors exhibiting large thermoelectric effects.
In the present study, this dissipation effect is introduced by $\varGamma_{\mathrm{ph}}(\omega)$, which is shown to be relatively small by the calibration to experimental $\kappa_{xx}$.
\par
The above discussion shows that being a dilute metal is advantageous for increasing $|\alpha^{\mathrm{PD}}_{xy}|$ and $|\kappa^{\mathrm{PD}}_{xy}|$, both in terms of electron and in terms of phonon.

    \paragraph{Conclusion.}
    \label{section:conclusion}
    We have obtained a very general and precise formula Eq. \eqref{eq:general_result} for Nernst and thermal Hall conductivities due to phonon drag effect in terms of the thermal Green's function, by combining linear response theory and the LK representation, and by precisely incorporating first-order perturbations of the magnetic field.
For \fix{various systems in which} the electron and phonon Green's functions are known, the conductivities can be calculated by this formula.
For metals with free electrons, the coefficients depend almost linearly on the effective mass.
In addition, it was shown that they peak at small Fermi energies of a few $\mathrm{meV}$.
The theory of the present letter is in very good agreement with the experiments on SrTiO$_{3-\delta}$ and explains why large coefficients are observed for dilute metals.
\fix{In future works, it is essential to incorporate the effect of vertex corrections and to compute the self-energy of phonons fully microscopically. However, these tasks are beyond the scope of this letter. The formula obtained in the present letter is expected to be a very essential guide in the search for materials with unusual thermoelectric coefficient behavior.}

    \paragraph{Acknowledgements.}
    J. Endo was supported by the Japan Society for the Promotion of Science and the Leading House ETH Zurich through the Program for Leading Graduate Schools (MERIT) and Young Researchers' Exchange Programme (No. JP\_EG\_special\_032023\_16).
This work is supported by Grants-in-Aid for Scientific Research from the Japan Society for the Promotion of Science (Nos. JP22KJ1055, JP22KK0228, and JP23K03274), and JST-Mirai Program Grant (No. JPMJMI19A1).

    \bibliographystyle{apsrev4-2}
    \bibliography{reference}
\fi

\ifshowsupp
    \widetext
    \clearpage
    \begin{center}
        \textbf{\large
            Supplementary Material
        }
    \end{center}
    \setcounter{section}{0}
    \setcounter{equation}{0}
    \setcounter{figure}{0}
    \setcounter{table}{0}
    \setcounter{page}{1}
    \makeatletter
    \renewcommand{\thesection}{S-\Roman{section}}
    \renewcommand{\theequation}{S\arabic{equation}}
    \renewcommand{\thefigure}{S\arabic{figure}}
    \renewcommand{\thetable}{S\arabic{table}}
    \label{section:supplementary}
    \section{Derivation of Thermal Current Operator of Acoustic Phonons and Electron-Phonon Interaction}
In this section, we discuss various aspects of the heat current of phonons.
We use a standard Hamiltonian density of acoustic phonons
\begin{equation}
    \hat{h}^{\mathrm{ph}}(\vr) = \frac{\hat{\vb*{P}}(\vr)^{2}}{2M} + \frac{Mc_{L}^{2}}{2}\sum_{\mu}\pqty{\pdv{\hat{\vb*{u}}(\vr)}{r_{\mu}}}^{2},
\end{equation}
where $\hat{\vb*{P}}(\vr)$ and $\hat{\vb*{u}}(\vr)$ are the momentum and the position of lattice displacement, and $M$ and $c_{L}$ are the mass of atoms and the sound velocity, respectively.
Assuming a longitudinal acoustic phonon, we rewrite $\hat{\vb*{u}}(\vr)$ and $\hat{\vb*{P}}(\vr)$ as
\begin{align}
    \hat{\vb*{u}}(\vr) & = \frac{1}{\sqrt{V}} \sum_{\vp} \mathrm{e}^{i\vp\cdot\vr} \vb*{e}_{\vp} \sqrt{\frac{\hbar}{2Mc_{L}\abs{\vp}}}~\pha[\vp],
    \\
    \hat{\vb*{P}}(\vr) & = \frac{i}{\sqrt{V}} \sum_{\vp} \mathrm{e}^{i\vp\cdot\vr} \vb*{e}_{\vp} \sqrt{\frac{\hbar Mc_{L}\abs{\vp}}{2}}~\phb[-\vp],
\end{align}
with
\begin{align}
    \pha[\vp] & = \hat{a}_{\vp} + \hat{a}^{\dagger}_{-\vp},  \\
    \phb[\vp] & = -\hat{a}_{-\vp} + \hat{a}^{\dagger}_{\vp}.
\end{align}
Here, $\vb*{e}_{\vp}$ is a unit phonon displacement vector and $\hat{a}_{\vp} (\hat{a}^{\dagger}_{\vp})$ is the annihilation (creation) operator of a phonon, which satisfies $\comm{\hat{a}_{\vp}}{\hat{a}_{\vp'}} = \comm{\hat{a}_{\vp}^{\dagger}}{\hat{a}_{\vp'}^{\dagger}}= 0$, and $\comm{\hat{a}_{\vp}}{\hat{a}_{\vp'}^{\dagger}} = \delta_{\vp,\vp'}$. As a result, $\comm{\pha[\vp]}{\pha[\vp']} = \comm{\phb[\vp]}{\phb[\vp']} = 0$ and $\comm{\pha[\vp]}{\phb[\vp']} = 2\delta_{\vp,\vp'}$ hold.
Then, the Fourier transform of the Hamiltonian density becomes
\begin{equation}
    \label{eq:Hamiltonian_density_of_phonon}
    \begin{split}
        \hat{h}^{\mathrm{ph}}_{\vq} & = \int \dd \vr ~ \mathrm{e}^{-i\vq\cdot\vr} \hat{h}^{\mathrm{ph}}(\vr)
        \\
                                    & = \frac{\hbar c_{L}}{4}\sum_{\vp} \abs{\vp} \pqty{\pha[\vp + \frac{\vq}{2}]\pha[-\vp + \frac{\vq}{2}] - \phb[\vp - \frac{\vq}{2}]\phb[-\vp - \frac{\vq}{2}]} + O\pqty{\abs{\vq}^{2}}.
    \end{split}
\end{equation}
\par
For phonons, the thermal current operator $\hat{\vb*{j}}^{2,\mathrm{ph}}(\vr)$ is equal to the energy current operator, which satisfies the continuity equation,
\begin{equation}
    \nabla \cdot \hat{\vb*{j}}^{2,\mathrm{ph}}(\vr) = \frac{i}{\hbar}\comm{\hat{h}^{\mathrm{ph}}(\vr)}{\ham^{\mathrm{ph}}},
\end{equation}
or equivalently,
\begin{equation}
    \vq \cdot \hat{\vb*{j}}^{2,\mathrm{ph}}_{\vq} = \frac{1}{\hbar} \comm{\hat{h}^{\mathrm{ph}}_{\vq}}{\hat{h}^{\mathrm{ph}}_{\vb*{0}}},
\end{equation}
where $\hat{\vb*{j}}^{2,\mathrm{ph}}_{\vq}$ is the Fourier transform of
$\hat{\vb*{j}}^{2,\mathrm{ph}}(\vr)$, defined by
$\hat{\vb*{j}}^{2,\mathrm{ph}}_{\vq}= \int \dd \vr ~ \mathrm{e}^{-i\vq\cdot\vr}
    \hat{\vb*{j}}^{2,\mathrm{ph}}(\vr)$.
Using Eq. \eqref{eq:Hamiltonian_density_of_phonon}, we obtain
\begin{equation}
    \frac{1}{\hbar} \comm{\hat{h}^{\mathrm{ph}}_{\vq}}{\hat{h}^{\mathrm{ph}}_{\vb*{0}}} = \vq \cdot \sum_{\vp} \frac{\hbar c_{L}^{2}}{2}\vp ~ \phb[\vp - \frac{\vq}{2}]\pha[\vp + \frac{\vq}{2}] + O\pqty{\abs{\vq}^{2}} .
\end{equation}
Therefore, the equation,
\begin{equation}
    \label{eq:j2_phonon_operator}
    \hat{\vb*{j}}^{2,\mathrm{ph}}_{\vq} = \sum_{\vp} \frac{\hbar c_{L}^{2}}{2}\vp ~ \phb[\vp - \frac{\vq}{2}]\pha[\vp + \frac{\vq} {2}] + O\pqty{\abs{\vq}} ,
\end{equation}
holds.
Note that there are several complicated terms in the order of $\abs{\vq}^{1}$ of this operator. However, as we show later, they do not contribute to the Nernst calculation.
\par
To calculate the transport coefficients, we define phonon Green's functions as
\begin{equation}
    \label{eq:supp_phonon_green}
    \begin{split}
        \phD{\vp}{\tau - \tau'} & = - \mean{\mathrm{T}_{\tau} \pha[\vp]\pqty{\tau}\pha[-\vp]\pqty{\tau'}},                                                                                                        \\
        \phC{\vp}{\tau - \tau'} & = - \mean{\mathrm{T}_{\tau} \pha[\vp]\pqty{\tau}\phb[\vp]\pqty{\tau'}},                                                                                                         \\
        \phD{\vp}{\iwm}         & = - \int_{0}^{1/\kbt} \dd \tau ~ \mathrm{e}^{\iwm\tau}\mean{\pqty{\hat{a}_{\vp}(\tau) + \hat{a}^{\dagger}_{-\vp}(\tau)}\pqty{\hat{a}_{-\vp}(0) + \hat{a}^{\dagger}_{\vp}(0)}},  \\
        \phC{\vp}{\iwm}         & = - \int_{0}^{1/\kbt} \dd \tau ~ \mathrm{e}^{\iwm\tau}\mean{\pqty{\hat{a}_{\vp}(\tau) + \hat{a}^{\dagger}_{-\vp}(\tau)}\pqty{-\hat{a}_{-\vp}(0) + \hat{a}^{\dagger}_{\vp}(0)}},
    \end{split}\end{equation}
where $\pha[\vp]\pqty{\tau}$ and $\phb[\vp]\pqty{\tau}$ are in Heisenberg representations of $\pha[\vp]$ and $\phb[\vp]$, respectively. Note that if there is no external field with nonzero wavenumbers, there is no need to use these Green's functions, but instead just use $\mathcal{O}$[15,29]. However, since we are considering perturbations with respect to vector potentials, we use the phonon Green's functions in
Eq. \eqref{eq:supp_phonon_green}.
\par
In addition, we discuss the interaction between electrons and phonons. Since we consider the longitudinal acoustic phonons, we assume that the Hamiltonian density of the electron-phonon interaction is given by
\begin{equation}
    \hat{h}^{\mathrm{e-p}}(\vr) = -g\hat{\rho}(\vr)\sum_{\mu}\pqty{\pdv{u_{\mu}(\vr)}{r_{\mu}}},
\end{equation}
where $\hat{\rho}(\vr)$ is the density operator of electrons, and $g$ is a constant. Therefore, the electron-phonon Hamiltonian becomes
\begin{equation}
    \begin{split}
        \ham^{\mathrm{e-p}} = & \int \dd \vr ~ \hat{h}^{\mathrm{e-p}}(\vr)                                                        \\
        =                     & \frac{1}{V} \sum_{\vp} g_{\vp} \hat{\vb*{c}}^{\dagger}_{\vk + \vp} \hat{\vb*{c}}_{\vk} \pha[\vp],
    \end{split}
\end{equation}
with $g_{\vp} = g \sqrt{\hbar\abs{\vp} / 2Mc_{L}}$.
In the presence of the electron-phonon interaction, there are also cross terms such as $\comm{\hat{h}^{\mathrm{e-p}}}{\ham^{\mathrm{ph}}}$ and $\comm{\hat{h}^{\mathrm{ph}}}{\ham^{\mathrm{e-p}}}$, which lead to the heat current originating from the electron-phonon interaction[29]. However, this heat current operator is not necessary in the discussion of the present letter.

\section{Electrons in Magnetic Field}
We use a Hamiltonian density of electrons
\begin{equation}
    \hat{h}^{\mathrm{ele}}(\vr) = \frac{1}{2m}\pqty{\hat{\vp} - e\vb*{A}(\vr)}^{2} + V(\vr),
\end{equation}
($e<0$ for electrons) and the vector potential has a form
\begin{equation}
    \vb*{A}(\vr) = - i\vb*{A}_{\vqz} \mathrm{e}^{i\vqz\cdot\vr},
\end{equation}
where $\vqz$ is a finite value. To guarantee the gauge invariance, the physical quantity should be proportional to
\begin{equation}
    q_{0,x}A_{\vqz, y} - q_{0,y}A_{\vqz, x}.
\end{equation}
At the end, we take the limit of $\vqz \to \vb*{0}$.
The Bloch wave functions are given by $\psi_{\ell \vk} (\vr) = \mathrm{e}^{i\vk\cdot\vr}u_{\ell\vk}(\vr)$, where $\ell$ represents the combined band and spin index and $u_{\ell\vk}(\vr)$ is a lattice-periodic function satisfying
\begin{equation}
    \ham_{\vk} u_{\ell\vk}(\vr) = \ep_{\ell\vk}u_{\ell\vk}(\vr).
\end{equation}
with $\ham_{\vk} = \mathrm{e}^{-i\vk\cdot\vr} \ham \mathrm{e}^{i\vk\cdot\vr}$.
In the Luttinger-Kohn (LK) representation[26,27], the electron operator can be written as
\begin{equation}
    \hat{\psi}(\vr) = \sum_{\ell,\vk}\mathrm{e}^{i\vk\cdot\vr} u_{\ell\vk_{0}}(\vr) \hat{\vb*{c}}_{\ell\vk},
\end{equation}
with some fixed momentum $\vk_{0}$. Using this representation and the continuity equation
\begin{equation}
    \nabla \cdot \hat{\vb*{j}}^{1,\mathrm{ele}} (\vr) = \frac{ie}{\hbar}\comm*{\rho(\vr)}{\ham^{\mathrm{\mathrm{ele}}}},
\end{equation}
we obtain the Fourier transform of the current density operator
\begin{align}
    \hat{\vb*{j}}^{1,\mathrm{ele}}_{\vq}      & = \hat{\vb*{j}}^{1,\mathrm{ele,para}}_{\vq} + \vb*{j}^{1,\mathrm{ele,dia}}_{\vq},                                                                             \\
    \hat{\vb*{j}}^{1,\mathrm{ele,para}}_{\vq} & = \frac{e}{\hbar}\sum_{\vk} \hat{\vb*{c}}^{\dagger}_{\vk - \frac{\vq}{2}} \tilde{\vb*{\gamma}}_{\vk} \hat{\vb*{c}}^{\phantom{\dagger}}_{\vk + \frac{\vq}{2}}, \\
    \hat{\vb*{j}}^{1,\mathrm{ele,dia}}_{\vq}  & = i \frac{e^{2}}{m} \vb*{A}_{\vqz} \rho_{\vq - \vqz},
\end{align}
where
\begin{align}
    \pqty{\tilde{\vb*{\gamma}}_{\vk}}_{\ell\ell'} = & \int \dd \vr ~ u_{\ell\vk_{0}}^{*}(\vr) \pdv{\ham_{\vk}}{\vk} u_{\ell'\vk_{0}}(\vr), \\
    \pdv{\ham_{\vk}}{\vk} =                         & \frac{\hbar^{2}}{m} \left( \vk - i\bm \nabla \right).
\end{align}
Under these assumptions, the first-order Hamiltonian with respect to the vector potential is equal to
\begin{equation}
    i\pqty{\hat{\vb*{j}}^{1,\mathrm{ele,para}}_{-\vqz}\cdot \vb*{A}_{\vqz}} = \frac{ie}{\hbar}\vb*{A}_{\vqz}\cdot\sum_{\vk} \hat{\vb*{c}}^{\dagger}_{\vk + \frac{\vqz}{2}} \tilde{\vb*{\gamma}}_{\vk} \hat{\vb*{c}}^{\phantom{\dagger}}_{\vk - \frac{\vqz}{2}}.
\end{equation}

\section{Derivation of $L^{21,\mathrm{PD}}_{xy}$}
The correlation function is defined as
\begin{equation}
    \Phi^{ij}_{\mu\nu}(\vq,\iwlam) = \frac{1}{V}\int_{0}^{\beta}\dd \tau ~ \mean{j^{i}_{\vq\mu}(\tau) j^{j}_{\vb*{0}\nu}(0)} \mathrm{e}^{\iwlam \tau}.
\end{equation}
According to the linear response theory, $L^{ij}_{\mu\nu}(\vq,\omega)$ is obtained from
the analytic continuation of $\Phi^{ij}_{\mu\nu}(\vq,\iwlam)$ as
\begin{equation}
    L^{ij}_{\mu\nu} (\vq, \omega) = \frac{\Phi^{ij}_{\mu\nu}\pqty{\vq,\iwlam\to\hbar(\omega + i\delta)} - \Phi^{ij}_{\mu\nu}\pqty{\vq, \hbar(0 + i\delta)}}{i(\omega + i\delta)}.
\end{equation}
\par
Let us calculate $L^{21,\mathrm{PD}}_{xy}$ due to the phonon drag effect.
Since the external magnetic field has a momentum $\vqz$,
the correlation function corresponding to the phonon drag consists of $\mean{j^{i,\mathrm{ph}}_{\vqz x} j^{j,\mathrm{ele}}_{\vb*{0} y}}$. For the first order of the magnetic field, $\Phi^{21,(1)}_{xy}(\vqz, \iwlam)$ is the sum of the Feynman diagrams in Fig. \ref{fig:Nernst_diagram}. Corresponding to Fig. \ref{fig:Nernst_diagram}(a)-(h), $\Phi^{21,(1)}_{xy}(\vqz, \iwlam)$ has the following eight contributions:
\begin{align}
    \label{eq:supp_correlation_a}
     & \vkbttwo \sumkpnm \coeff{1} \gpp \gpm\phC{\vpm}{\iwm}\phD{\vpp}{\iwmp} \notag                                                            \\ &\quad\times
    \Trb{\gre(\vkm,\ien)\gre(\vk-\vp,\ien-\iwm)\gre(\vkp,\ienp)} A_{\vqz,y} ,                                                                   \\
    \label{eq:supp_correlation_b}
     & \vkbttwo \sumkpnm \coeff{1} \gpp \gpm \phC{\vpm}{\iwm}\phD{\vpp}{\iwmp} \notag                                                           \\ &\quad\times
    \Trb{\gre(\vkm,\ien)\gre(\vk+\vp,\ienp+\iwm)\gre(\vkp,\ienp)} A_{\vqz,y} ,                                                                  \\
    \label{eq:supp_correlation_c}
     & \vkbttwo \sumkpnma \coeff{2} \gpp \gpm \phC{\vpm}{\iwm}\phD{\vpp}{\iwmp} \notag                                                          \\ &~\times
    \Trb{\gamma_{\vkp,y}\gre(\vkp,\ien)\gamma_{\vk,\alpha}\gre(\vkm,\ien)\gre(\vk-\vp,\ien-\iwm)\gre(\vkp,\ienp)} A_{\vqz,\alpha} ,             \\
    \label{eq:supp_correlation_d}
     & \vkbttwo \sumkpnma \coeff{2} \gpp \gpm \phC{\vpm}{\iwm}\phD{\vpp}{\iwmp} \notag                                                          \\ &\quad\times
    \Trb{\gamma_{\vkp,y}\gre(\vkp,\ien)\gamma_{\vk,\alpha}\gre(\vkm,\ien)\gre(\vk+\vp,\ienp+\iwm)\gre(\vkp,\ienp)} A_{\vqz,\alpha} ,            \\
    \label{eq:supp_correlation_e}
     & \vkbttwo \sumkpnma \coeff{2} \gpp \gpm \phC{\vpm}{\iwm}\phD{\vpp}{\iwmp} \notag                                                          \\ &\quad\times
    \Trb{\gamma_{\vkm,y}\gre(\vkm,\ien)\gre(\vk-\vp,\ien-\iwm)\gre(\vkp,\ienp)\gamma_{\vk,\alpha}\gre(\vkm,\ienp)} A_{\vqz,\alpha} ,            \\
    \label{eq:supp_correlation_f}
     & \vkbttwo \sumkpnma \coeff{2} \gpp \gpm \phC{\vpm}{\iwm}\phD{\vpp}{\iwmp} \notag                                                          \\ &\quad\times
    \Trb{\gamma_{\vkm,y}\gre(\vkm,\ien)\gre(\vk+\vp,\ienp+\iwm)\gre(\vkp,\ienp)\gamma_{\vk,\alpha}\gre(\vkm,\ienp)} A_{\vqz,\alpha} ,           \\
    \label{eq:supp_correlation_g}
     & \vkbttwo \sumkpnma \coeff{2} \gpp \gpm \phC{\vpm}{\iwm}\phD{\vpp}{\iwmp} \notag                                                          \\ &\quad\times
    \Trb{\gamma_{\vk,y}\gre(\vk,\ien)\gre(\vkp-\vp,\ien -\iwm)\gamma_{\vk-\vp,\alpha}\gre(\vkm-\vp,\ien-\iwm)\gre(\vk,\ienp)} A_{\vqz,\alpha} , \\
    \label{eq:supp_correlation_h}
     & \vkbttwo \sumkpnma \coeff{2} \gpp \gpm \phC{\vpm}{\iwm}\phD{\vpp}{\iwmp} \notag                                                          \\ &\quad\times
    \Trb{\gamma_{\vk,y}\gre(\vk,\ien)\gre(\vkp+\vp,\ienp+\iwm)\gamma_{\vk+\vp,\alpha}\gre(\vkm+\vp,\ienp+\iwm)\gre(\vk,\ienp)} A_{\vqz,\alpha}.
\end{align}
Here $\vb*{k}_\pm = \vb*{k}\pm \vqz/2, \vb*{p}_\pm = \vb*{p}\pm \vqz/2,
    \ienp = \ien + i\omega_\lambda, \iwmp = \iwm + i\omega_\lambda$, and
$\mathrm{Tr}$ represents the trace over the band index $\ell$ including the spin degrees of freedom. First, we extract the first-order with respect to $\vqz$ to confirm the gauge invariance as discussed above.
\par
It is easy to see that the phonon part $[\gpp \gpm \phC{\vpm}{\iwm}\phD{\vpp}{\iwmp}]$
is common in all the contributions.
Let us study its first-order contribution with respect to $\abs{\vqz}$. In this case,
the electron part should be calculated with $\vqz=\vb*{0}$.
For example, the electron part ($\sum_{\vb*{k}}$ Tr $[\cdots ]$) with $\vqz=\vb*{0}$ in
Eq. \eqref{eq:supp_correlation_g}
can be rewritten as
\begin{equation}
    \begin{split}
         & \quad \sum_{\vk}\Trb{\gamma_{\vk,y}\gre(\vk,\ien)\gre(\vk-\vp,\ien -\iwm)\gamma_{\vk-\vp,\alpha}\gre(\vk-\vp,\ien-\iwm)\gre(\vk,\ienp)} \\
         & = \sum_{\vk}\Trb{\gamma_{\vk,y}\gre(\vk,\ien)\pqty{\pdv{k_{\alpha}}\gre(\vk-\vp,\ien-\iwm)}\gre(\vk,\ienp)}                             \\
         & = -\sum_{\vk}\mathrm{Tr}\left[ \pqty{\pdv{k_{\alpha}}\gamma_{\vk,y}}\gre(\vk,\ien)\gre(\vk-\vp,\ien-\iwm)\gre(\vk,\ienp) \right.        \\
         & \qquad + \gamma_{\vk,y}\pqty{\pdv{k_{\alpha}}\gre(\vk,\ien)}\gre(\vk-\vp,\ien-\iwm)\gre(\vk,\ienp)                                      \\
         & \qquad \left. + \gamma_{\vk,y}\gre(\vk,\ien)\gre(\vk-\vp,\ien-\iwm)\pqty{\pdv{k_{\alpha}}\gre(\vk,\ienp)} \right],
    \end{split}
\end{equation}
where we have used the integration by parts and
$\partial \gre/\partial k_\alpha = \gre \gamma_\alpha \gre$.
This relation indicates that the sum of Eqs. \eqref{eq:supp_correlation_a}, \eqref{eq:supp_correlation_c}, \eqref{eq:supp_correlation_e}, and \eqref{eq:supp_correlation_g} cancel out when we take the first-order contribution with respect to $\vqz$ in the phonon part.
The similar cancellation occurs in Eqs. \eqref{eq:supp_correlation_b}, \eqref{eq:supp_correlation_d}, \eqref{eq:supp_correlation_f}, and \eqref{eq:supp_correlation_h}, which can be obtained by replacing
$\vb*{p} \rightarrow -\vb*{p}$ and $\iwm \rightarrow -\iwm-i\omega_\lambda$ in
Eqs. \eqref{eq:supp_correlation_a}, \eqref{eq:supp_correlation_c}, \eqref{eq:supp_correlation_e}, and \eqref{eq:supp_correlation_g}.
These are the reason why we do not need the explicit form in the order of $O(|\vq|)$ in Eq. \eqref{eq:j2_phonon_operator}.
\par
Therefore, we use the 0-th order contributions of the phonon part, and evaluate the first-order contributions of the electron part with respect to $\vqz$.
By expanding the electron part, rearranging the several terms, and using $q_{0, x}A_{\vqz, y} - q_{0, y}A_{\vqz, x} \to B_{z}$, we obtain
\begin{equation} \label{eq:supp_general_result}
    \begin{split}
        \Phi_{xy}^{21,\mathrm{PD}}(\vqz\rightarrow \vb*{0}, \iwlam)
        = & \vkbttwo \sumkpnm \coeffb{1} \gp^{2} \phC{\vp}{\iwm}\phD{\vp}{\iwmp}       \\ &\quad\times
        \Trb{\gre\gre'\grep\gamma_{\vk,x}\grep - \gre\gamma_{\vk,x}\gre\gre'\grep +
        (\vb*{p} \rightarrow -\vb*{p}, \iwm \rightarrow -\iwm-i\omega_\lambda) }       \\
          & + \vkbttwo \sumkpnm \coeffb{2} \gp^{2} \phC{\vp}{\iwm}\phD{\vp}{\iwmp}     \\ &\quad\times
        \Trb{( \gamma_{\vk,y}\gre\gamma_{\vk,x}\gre\gamma_{\vk,y}\gre\gre'\grep
        - \gamma_{\vk,y}\gre\gamma_{\vk,y}\gre\gamma_{\vk,x}\gre\gre'\grep \right.     \\ &\qquad\qquad+
            \gamma_{\vk,y}\gre\gre'\grep\gamma_{\vk,x}\grep\gamma_{\vk,y}\grep
        - \gamma_{\vk,y}\gre\gre'\grep\gamma_{\vk,y}\grep\gamma_{\vk,x}\grep           \\
          & \qquad\qquad+
            \gamma_{\vk,y}\gre\gamma_{\vk,y}\gre\gre'\grep\gamma_{\vk,x}\grep
        - \gamma_{\vk,y}\gre\gamma_{\vk,x}\gre\gre'\grep\gamma_{\vk,y}\grep            \\ &\qquad\qquad+
            \gamma_{\vk,y}\gre\gre'\gamma_{\vk-\vp,x}\gre'\gamma_{\vk-\vp,y}\gre'\grep
        - \gamma_{\vk,y}\gre\gre'\gamma_{\vk-\vp,y}\gre'\gamma_{\vk-\vp,x}\gre'\grep ) \\
          & \qquad\qquad
            +  (\vb*{p} \rightarrow -\vb*{p}, \iwm \rightarrow -\iwm-i\omega_\lambda)
            \left. }.
    \end{split}
\end{equation}
Here, $\gre = \gre(\vk,\ien)$, $\grep = \gre(\vk,\ienp)$, $\gre' = \gre(\vk-\vp,\ien-\iwm)$,
and $(\vb*{p} \rightarrow -\vb*{p}, \iwm \rightarrow -\iwm-i\omega_\lambda)$ means the
contributions which has replacements  $\vb*{p} \rightarrow -\vb*{p}$ and
$\iwm \rightarrow -\iwm-i\omega_\lambda$ in the preceding terms.
\par
For $\Phi^{22}$, just add $\pqty{\ep_{\vk} - \mu}/e$ as a factor. This corresponds to changing the electron current to thermal current.
\par
\begin{figure}[htbp]
    \begin{center}
        \includegraphics[keepaspectratio, width=16cm]{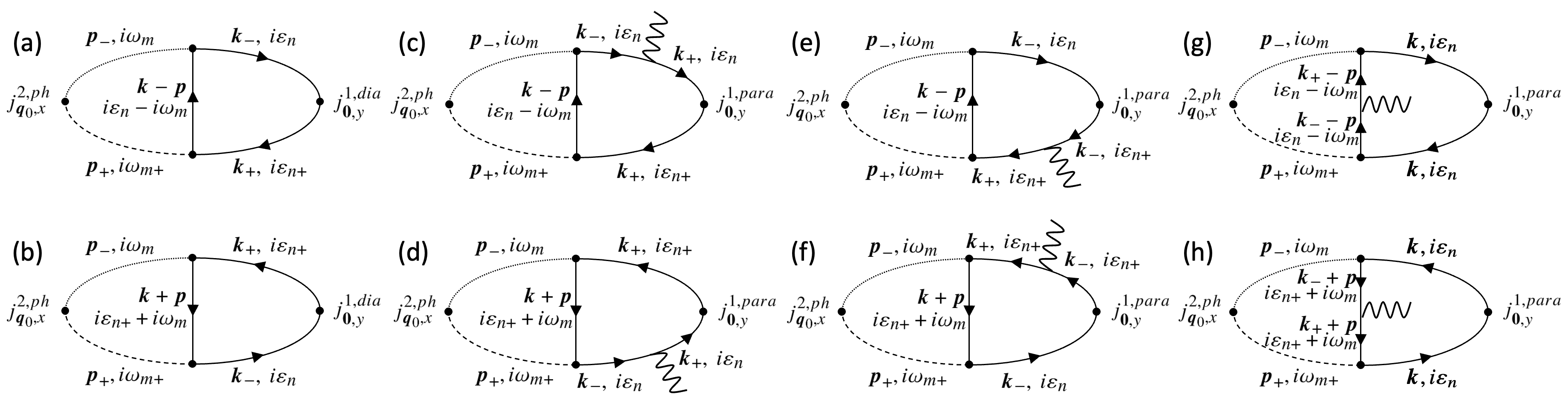}
        \caption{The Feynman diagrams corresponding to $\Phi^{21}_{xy}$ of the phonon drag effect.}
        \label{fig:Nernst_diagram}
    \end{center}
\end{figure}

\section{Derivation of Eq. (13)}
In this section, we calculate $L^{21,\mathrm{PD}}_{xy}$ in the case when the electron system is described as a single-band free electron with dispersion $\ep_{\vk} = \hbar^{2}\abs{\vk}^{2}/2m^{*}$.
In the case of free electrons, the current and heat current operators are given by
\begin{align}
    \hat{\vb*{j}}^{1,\mathrm{ele}}_{\vq,\mu} & = e\sum_{\vk} v_{\vk,\mu} \hat{c}^{\dagger}_{\vk - \frac{\vq}{2}} \hat{c}^{\phantom{\dagger}}_{\vk + \frac{\vq}{2}},                  \\
    \hat{\vb*{j}}^{2,\mathrm{ele}}_{\vq,\mu} & = \sum_{\vk} v_{\vk,\mu} (\ep_{\vk} - \mu) \hat{c}^{\dagger}_{\vk - \frac{\vq}{2}} \hat{c}^{\phantom{\dagger}}_{\vk + \frac{\vq}{2}},
\end{align}
where $v_{\vk,\mu} = \hbar k_{\mu} / m^{*}$ is the velocity of a electron.
In this case, we can easily see that the second term of Eq. \eqref{eq:supp_general_result} vanishes due to the symmetry.
Therefore, we obtain
\begin{equation}
    \begin{split}
        \Phi_{xy}^{21,\mathrm{PD}}(i\omega_{\lambda})
        = & -\frac{k_{\rm B}T}{V^2} \sum_{\vb*{k}, \vb*{p}, m} 
        \frac{ie^{2}B_{z}\hbar^3 c_{L}^{2} p_x k_x}{2(m^{*})^2}
        \gp^{2} \phC{\vp}{\iwm}\phD{\vp}{\iwmp}                \\ &\quad\times
        \left[  F(\vk, \vp, \iwm , i\omega_{\lambda}) +  F(\vk, -\vp, -\iwm -i\omega_\lambda, i\omega_{\lambda}) \right],
    \end{split}
\end{equation}
where the spin summation has been taken into account and
\begin{equation}
    F(\vk, \vp, \iwm , i\omega_{\lambda}) = -k_{B}T \sum_{n}\pqty{\gre\gre'\grep\grep-\gre\gre\gre'\grep}.
\end{equation}
Since we consider the free electrons and acoustic phonons, we use
\begin{equation}
    \gre(\vk,\ien) = \frac{1}{\ien - \ep_{\vk} + \mu + i\varGamma~\mathrm{sign}(\ep_{n})},
\end{equation}
and
\begin{equation}
    \phD{\vq}{\iwm} = \frac{1}{\iwm - \hbar \omega_{\vq} + i\varGamma_{\mathrm{ph}}(\iwm)~\mathrm{sign}(\omega_{m})} - \frac{1}{\iwm + \hbar \omega_{\vq} + i\varGamma_{\mathrm{ph}}(\iwm)~\mathrm{sign}(\omega_{m})},
\end{equation}
\begin{equation}
    \phC{\vq}{\iwm} = \frac{1}{\iwm - \hbar \omega_{\vq} + i\varGamma_{\mathrm{ph}}(\iwm)~\mathrm{sign}(\omega_{m})} + \frac{1}{\iwm + \hbar \omega_{\vq} + i\varGamma_{\mathrm{ph}}(\iwm)~\mathrm{sign}(\omega_{m})},
\end{equation}
where $\hbar \omega_{\vq} = \hbar c_{L} \abs{\vq}$ is the energy of a phonon, and
we consider the $\iwm$-dependence of the phonon relaxation rate later.
\par
To extract the dominant contribution assuming that the phonon relaxation rate is small
($\varGamma_{\mathrm{ph}} \ll E_{\mathrm F}$), we consider the situation where
$-\omega_\lambda < \omega_m <0$, as was assumed in the phonon-drag problems[15, 29].
In this region, we obtain
\begin{equation}
    \begin{split}
         & F(\vk, \vp, \iwm , i\omega_{\lambda})
        =\int_{-\infty}^{\infty}\frac{\dd \ep}{2\pi i} f(\ep)                                                          \\ &\times \biggl[
        \left\{  [G^R(\vk,\ep+i\omega_\lambda)]^2 \pqty{G^{R}(\vk,\ep) - G^{A}(\vk, \ep)}
        - G^R(\vk,\ep+i\omega_\lambda) \pqty{[G^{R}(\vk,\ep)]^2 - [G^{A}(\vk, \ep)]^2} \right\} G^R(\vk-\vp, \ep-\iwm) \\
         & + \left\{ [G^{R}(\vk,\ep+\iwm+i\omega_\lambda)]^2 G^{A}(\vk, \ep+\iwm)
        -G^{R}(\vk,\ep+\iwm+i\omega_\lambda) [G^{A}(\vk, \ep+\iwm)]^2 \right\}
        \pqty{G^{R}(\vk-\vp,\ep) - G^{A}(\vk-\vp, \ep)}                                                                \\
         & + \left\{ \pqty{[G^{R}(\vk,\ep)]^2 - [G^{A}(\vk, \ep)]^2} G^A(\vk,\ep-i\omega_\lambda)
        -  \pqty{G^{R}(\vk,\ep) - G^{A}(\vk, \ep)} [G^A(\vk,\ep-i\omega_\lambda)]^2
        \right\} G^A(\vk-\vp, \ep-\iwm-i\omega_\lambda) \biggl].
    \end{split}
\end{equation}
Then, we take the $\iwm$-summation in the region $-\omega_\lambda < \omega_m <0$ to obtain
\begin{equation}
    \begin{split}
        \Phi_{xy}^{21,\mathrm{PD}}(i\omega_{\lambda})
        = & \frac{1}{V^2} \sum_{\vb*{k}, \vb*{p}}\int_{-\infty}^\infty \frac{d z}{2\pi i} n(z)
        \frac{ie^{2}B_{z}\hbar^3 c_{L}^{2} p_x k_x}{2(m^{*})^2} \gp^{2}                                                     \\
          & \times \biggl\{ C^A(\vp, z) D^R(\vp, z+i\omega_\lambda)
        \left[  F(\vk, \vp, z , i\omega_{\lambda}) +  F(\vk, -\vp, -z -i\omega_\lambda, i\omega_{\lambda}) \right]          \\
          & - C^A(\vp, z-i\omega_\lambda) D^R(\vp, z)
        \left[  F(\vk, \vp, z-i\omega_\lambda , i\omega_{\lambda}) +  F(\vk, -\vp, -z, i\omega_{\lambda}) \right] \biggr\}. \\
    \end{split}
\end{equation}
Finally, using the analytic continuation $i\omega_{\lambda} \to \hbar\omega + i\delta$, we obtain
\begin{equation}
    \begin{split}
        L_{xy}^{21,\mathrm{PD}} = & \lim_{\omega \to 0}\frac{\Phi_{xy}^{21,\mathrm{PD}}(i\omega_{\lambda} \to \hbar\omega + i\delta)}{i(\omega+i\delta)} \\
        =                         & -\frac{1}{V^2} \sum_{\vb*{k}, \vb*{p}}\int_{-\infty}^\infty \frac{d z}{2\pi i}
        \frac{\partial n(z)}{\partial z}
        \frac{e^{2}B_{z}\hbar^4 c_{L}^{2} p_x k_x}{2(m^{*})^2} \gp^{2}                                                                                   \\
                                  & \times C^A(\vp, z) D^R(\vp, z)
        \left[  F(\vk, \vp, z , i\delta) +  F(\vk, -\vp, -z, i\delta) \right],
    \end{split}
\end{equation}
where the electron part becomes
\begin{equation}
    \begin{split}
          & F(\vk, \vp, z, i\delta)                                                                                        \\
        = & \int_{-\infty}^{\infty}\frac{\dd \ep}{2\pi i} f(\ep) \biggl[
        -G^R(\vk,\ep) G^A(\vk,\ep) \pqty{G^{R}(\vk,\ep) - G^{A}(\vk, \ep)} \pqty{G^R(\vk-\vp, \ep-z) -G^A(\vk-\vp, \ep-z)} \\
          & + G^{R}(\vk,\ep+z)G^{A}(\vk, \ep+z) \pqty{G^{R}(\vk,\ep+z) -G^{A}(\vk, \ep+z)}
        \pqty{G^{R}(\vk-\vp,\ep) - G^{A}(\vk-\vp, \ep)} \biggl].                                                           \\
        = & \int_{-\infty}^{\infty}\frac{\dd \ep}{2\pi i} \pqty{f(\ep - z) - f(\ep)} A(\ep,z)
    \end{split}
\end{equation}
with
\begin{equation}
    A(\ep,z) = G^{R}(\vk,\ep)G^{A}(\vk, \ep)\pqty{G^{R}(\vk,\ep) - G^{A}(\vk, \ep)} \pqty{G^{R}(\vk-\vp,\ep-z) - G^{A}(\vk-\vp, \ep-z)}.
\end{equation}
\par
In the following, we evaluate $L_{xy}^{21,\mathrm{PD}}$ assuming the small relaxation rates.
For the phonon part, we use
\begin{equation}
    C^{A}(\vp,z)D^{R}(\vp,z) \simeq \frac{\pi}{\varGamma_{\mathrm{ph}}(z)}\pqty{\delta(z - \omega) - \delta(z + \omega)},
\end{equation}
and for the electron part,
\begin{align}
    G^{R}(\vk,\ep)G^{A}(\vk, \ep)\pqty{G^{R}(\vk,\ep) - G^{A}(\vk, \ep)} \simeq -\frac{i\pi}{\varGamma^{2}}\delta(\ep - \epk), \\
    G^{R}(\vk-\vp,\ep-z) - G^{A}(\vk-\vp, \ep-z)=-2\pi i \delta(\ep-z-\ep_{\vk-\vp}).
\end{align}
Putting $g_{\vp}^{2} = g'^{2} |\vp|$, we obtain
\begin{equation}
    \begin{split}
        L_{xy}^{21,\mathrm{PD}}
        = & -\frac{\pi e^{2}B_{z}\hbar^4 c_{L}^{2} g'^2}{2(m^{*})^2} \frac{1}{V^2} \sum_{\vb*{k}, \vb*{p}}
        \frac{|\vp| p_x k_x}{\varGamma^2\varGamma_{\mathrm{ph}}(\omega)}
        \frac{\partial n(\omega)}{\partial \omega}                                                                                         \\
          & \times (f(\ep_{\vk-\vp})-f(\ep_{\vk})) [ \delta(\ep_{\vk} -\omega-\ep_{\vk-\vp}) -  \delta(\ep_{\vk} + \omega-\ep_{\vk-\vp})],
        \label{eq:L21finalXX}
    \end{split}
\end{equation}
where we have assumed that $\varGamma_{\mathrm{ph}}(z)$ is an even function of $z$.
Since $\ep_{\vk}$ is an even function of $\vk$, the change of variable
$\vk \rightarrow -\vk+\vp$ leads to the relations
$\ep_{\vk} \rightarrow \ep_{-\vk+\vp}=\ep_{\vk-\vp}$ and
$\ep_{\vk-\vp} \rightarrow \ep_{-\vk}=\ep_{\vk}$. Using this change of variable,
we can rewrite the second delta function in Eq.~(\ref{eq:L21finalXX}).
Then, we obtain
\begin{equation}
    \begin{split}
        \frac{L_{xy}^{21,\mathrm{PD}}}{B_{z}}
        = & -\frac{\pi e^{2}\hbar^4 c_{L}^{2} g'^2}{2(m^{*})^2} \frac{1}{V^2} \sum_{\vb*{k}, \vb*{p}}
        \frac{\abs{\vp} p_x^2}{\varGamma^2\varGamma_{\mathrm{ph}}(\omega)}
        \frac{\partial n(\omega)}{\partial \omega}
        (f(\ep_{\vk-\vp})-f(\ep_{\vk})) \delta(\ep_{\vk} -\omega-\ep_{\vk-\vp})                                                                                                                                                                     \\
          & = -\frac{e^{2}g'^{2}}{48\pi^{3}\hbar^{5}c_{L}^3}\int_{m^*c_L^2/2}^{\infty}\dd \ep \int_{0}^{\alpha} \dd \omega ~ \frac{\omega^{4}}{\varGamma^{2}\varGamma_{\mathrm{ph}}(\omega)}\pqty{f(\ep - \omega) - f(\ep)}\pdv{n(\omega)}{\omega},
        \label{eq:L21finalXX2}
    \end{split}
\end{equation}
with $\alpha = 2\sqrt{2mc_L^{2}\ep} - 2mc_L^{2}$, and we have put $\hbar\omega_{\vq} = \omega$.
\par
Similarly, we obtain
\begin{equation}
    \frac{L_{xy}^{22,\mathrm{PD}}}{B_{z}}
    = -\frac{e^{2}g'^{2}}{48\pi^{3}\hbar^{5}c_{L}^3}\int_{m^*c_L^2/2}^{\infty}\dd \ep \int_{0}^{\alpha} \dd \omega ~ \frac{\pqty{\ep - \mu}\omega^{4}}{\varGamma^{2}\varGamma_{\mathrm{ph}}(\omega)}\pqty{f(\ep - \omega) - f(\ep)}\pdv{n(\omega)}{\omega}.
\end{equation}

\section{Calculation of Chemical potential}
The density of states of free electrons in a 3D system is
\begin{equation}
    D(\ep) = \frac{V}{2\pi^{2}} \pqty{\frac{2m^{*}}{\hbar^{2}}}^{3/2} \sqrt{\ep}.
\end{equation}
The chemical potential is determined by the self-consistent equation
\begin{equation}
    nV = \int \dd \ep ~ f(\ep)D(\ep).
\end{equation}
We show the density of states and the chemical potential in Figure \ref{fig:ChemicalPotential}.

\begin{figure}[tbp]
    \begin{center}
        \includegraphics[keepaspectratio, width=0.5\linewidth]{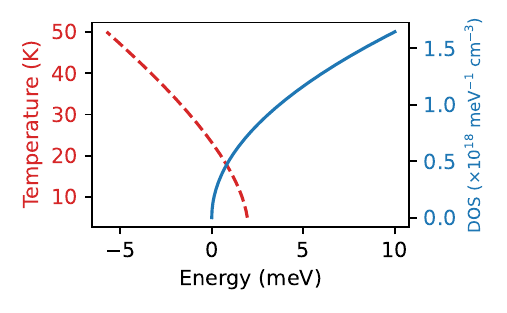}
        \caption{Temperature dependence of chemical potential (red
            dashed line) and density of states (blue solid line).}
        \label{fig:ChemicalPotential}
    \end{center}
\end{figure}

\section{Calculation of Seebeck Coefficient}
For free electrons, the correlation functions can be written as
\begin{align}
    \Phi^{11}_{xx}(\vb*{0}, \iwlam) & = -\frac{\kbt}{V}\sum_{\vk, n} \pqty{\frac{e\hbar k_{x}}{m^{*}}}^{2} \gre(\vk, \ienp)\gre(\vk, \ien),
\end{align}
without any perturbations.
Using this, we can calculate the conductivity as
\begin{equation}
    L^{11}_{xx} = \frac{e^{2}\sqrt{m^{*}}}{6\pi^{2}\hbar^{2}} \int_{-\infty}^{\infty} \dd \ep ~ \pqty{-f'(\ep)} \frac{\varGamma^{2}}{\pqty{\sqrt{\ep^{2}+\varGamma^{2}} - \ep}^{3/2}},
\end{equation}
with small $\varGamma$. Using the Sommerfeld-Bethe relation, we obtain
\begin{equation}
    L^{12, \mathrm{ele}}_{xx} = \frac{e\sqrt{m^{*}}}{6\pi^{2}\hbar^{2}}  \int_{-\infty}^{\infty} \dd \ep ~ \pqty{-f'(\ep)} (\ep - \mu) \frac{\varGamma^{2}}{\pqty{\sqrt{\ep^{2}+\varGamma^{2}} - \ep}^{3/2}}.
    \label{eq:L11XX}
\end{equation}
\par
For the phonon drag, Ogata shows[29]
\begin{equation}
    L_{xx}^{12,\mathrm{PD}} = \frac{1}{2V^{2}}\sum_{\vk,\vq} \frac{ec_{L}^{2}\pi \hbar^{2}g_{\vq}^{2}q_{x}v_{\vk,x}}{\varGamma\varGamma_{\mathrm{ph}}}n'(\hbar\omega_{\vq}) \pqty{f(\ep_{\vk}) - f(\ep_{\vk - \vq})}\pqty{\delta(\ep_{\vk} - \ep_{\vk - \vq} - \hbar\omega_{\vq}) - \delta(\ep_{\vk} - \ep_{\vk - \vq} + \hbar\omega_{\vq})},
    \label{eq:L12ELEXX}
\end{equation}
and taking the summation over the wavenumbers, we obtain
\begin{equation}
    L_{xx}^{12,\mathrm{PD}} = -\frac{eg'^{2}m^{*}}{48\pi^{3}\hbar^{6}c_{L}^{3}}\int_{m^*c_L^2/2}^{\infty}\dd \ep \int_{0}^{\alpha} \dd \omega ~ \frac{\omega^{4}}{\varGamma\varGamma_{\mathrm{ph}}(\omega)}\pqty{f(\ep - \omega) - f(\ep)}\pdv{n(\omega)}{\omega}.
    \label{eq:L12PDXX}
\end{equation}

\section{Calibration of $\varGamma_{\mathrm{ph}}$}
To compare our theory with experimental data, we assume the following form for the spectrum of the scattering rate of the phonon:
\begin{equation}
    \varGamma_{\mathrm{ph}}(\omega) = \varGamma_{\mathrm{ph}}^{(0)} \pqty{1 + A\omega^{4}+B\kbt\omega^{2}\mathrm{e}^{-\hbar\omega_{D}/3\kbt}},
\end{equation}
where $\omega_{D}$ is the Debye frequency.
The thermal conductivity is calculated as ( P. Carruthers, Rev. Mod. Phys. \textbf{33}, 92 (1961) )
\begin{equation}
    \kappa_{xx} = \frac{1}{4\pi^{2}\hbar^{2}\tilde{c}T} \int_{0}^{\hbar\omega_{D}} \dd\omega ~ (-n'(\omega))\frac{\omega^{4}}{\varGamma_{\mathrm{ph}}(\omega)},
\end{equation}
where $\tilde{c}$ is the mean velocity of acoustic phonon determined as $1/c_{L} + 2/c_{T} = 3 / \tilde{c}$ where $c_{T}$ is the velocity of transverse acoustic phonon.
In addition to the parameters in the text, we use $c_{T} = 4150~\mathrm{m/s}$[32] and $\hbar \omega_{D} / k_{B} = 400~\mathrm{K}$ as a typical value. The coefficients are determined as $\varGamma_{\mathrm{ph}}^{(0)} \simeq 4.7 \times 10^{-7}~\mathrm{eV}$, $A \simeq 1.2 \times 10^{7}~\mathrm{eV^{-4}}$, and $B \simeq 1.7 \times 10^{9}~\mathrm{eV^{-2}}$, so as to reproduce the experimental data of longitudinal thermal conductivity $\kappa_{xx}$ shown in Ref. [20].
A comparison of thermal conductivity with the fitting function and experimental data is shown in Figure \ref{fig:ThermalConductivity}.

\par
\begin{figure}[tbp]
    \begin{center}
        \includegraphics[keepaspectratio, width=0.5\linewidth]{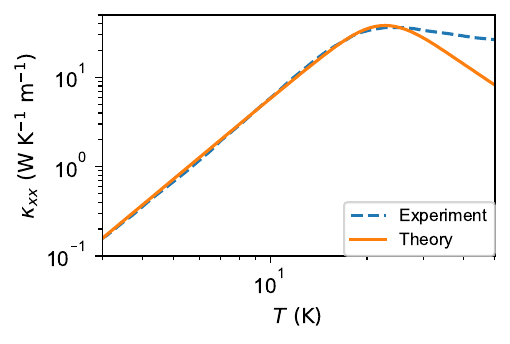}
        \caption{Temperature dependence of thermal conductivity in our theory (orange line) and experimental data[20] (blue dotted line). The calculated thermal conductivity is computed after fitting the parameters, assuming the phonon scattering rate as Eq. (14). Although our calculations only consider acoustic phonons, optical phonons also contribute to thermal conductivity at high temperatures. The discrepancy between experiment and theory at high temperatures may be due to such reason.}
        \label{fig:ThermalConductivity}
    \end{center}
\end{figure}

\section{Dependence on Effective Mass $m^{*}$}
Assuming $\varGamma_{\mathrm{ph}}$ is constant, we calculate dependence of thermoelectric coefficients on the effective mass using Eq. (13).
Figure \ref{fig:Mass} shows $L^{21}_{xy}/L^{21}_{xy}(m^{*}c_{L}^{2}/ k_{B}T = 1)$ and $L^{22}_{xy}/L^{22}_{xy}(m^{*}c_{L}^{2}/ k_{B}T = 1)$.
We can see that the coefficients increase with a nearly linear dependence until the effective mass reaches about $2k_{B}T$.
\begin{figure}[tbp]
    \begin{minipage}[b]{0.48\linewidth}
        \centering
        \includegraphics[keepaspectratio, width=\linewidth]{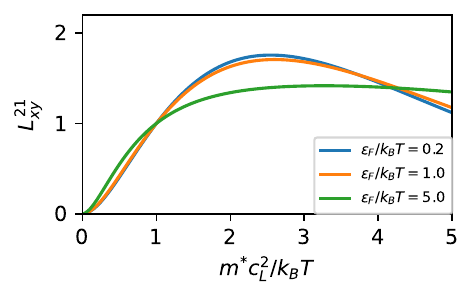}
    \end{minipage}
    \begin{minipage}[b]{0.48\linewidth}
        \centering
        \includegraphics[keepaspectratio, width=\linewidth]{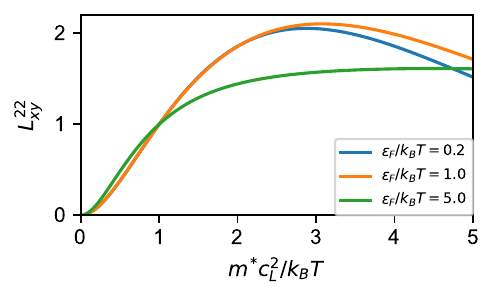}
    \end{minipage}
    \caption{The ratio of the thermoelectric coefficients to that of $m^{*}c_{L}^{2}/ k_{B}T = 1$.
    In the region where the effective mass is sufficiently small, the thermoelectric coefficients are found to be approximately proportional to the effective mass.}
    \label{fig:Mass}
\end{figure}

\section{Dependence on Fermi Energy $\ep_{F}$}
To confirm that small Fermi energy causes large thermoelectric coefficient, the Fermi energy dependence of the thermoelectric coefficient is shown.
Here, $\varGamma \propto k_{F}$ is assumed and other constants are the values of SrTiO$_{3-\delta}$ in the text.
The results in Fig. \ref{fig:Fermi} are obtained using Eq. (13).
It can be seen that the coefficients $\alpha_{xy}^{\mathrm{PD}}$ and $\kappa'^{\mathrm{PD}}_{xy}$ have peak at around $\ep_{F} \sim 10~\mathrm{meV}$.
\begin{figure}[tbp]
    \begin{minipage}[b]{0.48\linewidth}
        \centering
        \includegraphics[keepaspectratio, width=\linewidth]{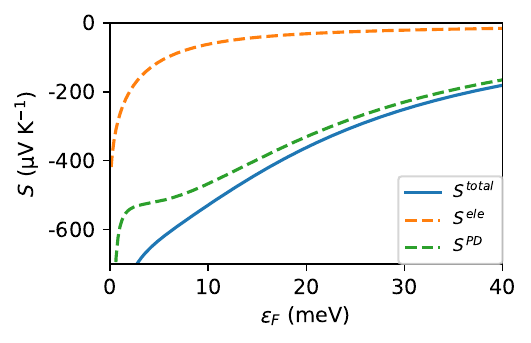}
    \end{minipage}
    \\
    \begin{minipage}[b]{0.48\linewidth}
        \centering
        \includegraphics[keepaspectratio, width=\linewidth]{image/Fermi_L21.pdf}
    \end{minipage}
    \begin{minipage}[b]{0.48\linewidth}
        \centering
        \includegraphics[keepaspectratio, width=\linewidth]{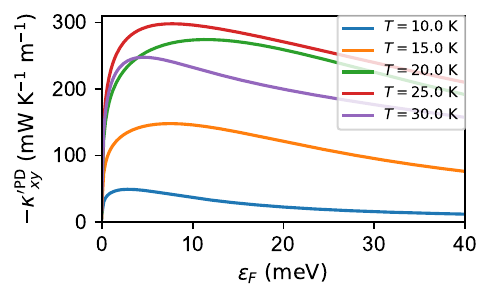}
    \end{minipage}
    \caption{Fermi energy dependence of the thermoelectric coefficients.
        The Seebeck coefficient is the value at $20~\mathrm{K}$.
        The values are obtained by hypothetically varying the Fermi energy of SrTiO$_{3-\delta}$.
        In the calculation, the $\varGamma$ is assumed to be proportional to the Fermi wavenumber $k_{F}$, and the other constants other than the chemical potential $\mu$ are assumed to be independent of the Fermi energy.}
    \label{fig:Fermi}
\end{figure}

\fi

\end{document}